\documentclass[11pt]{article}
\newcommand{\ket}[1]{\left | #1 \right \rangle}
\newcommand{\bra}[1]{\left \langle #1 \right |}

\def\openone{\leavevmode\hbox{\small1\kern-3.8pt\normalsize1}}
\def\RR{{\rm I\kern-.2emR}}
\def\tr{{\rm tr}\; }

\def\cg{{\cal G}}

\def\cp{{\cal P}}

\def\ca{{\cal A}}

\def\cs{{\cal S}}
\def\cd{{\cal D}}
\def\cq{{\cal Q}}
\def\ck{{\cal K}}

\def\poly{{\rm poly}}

\newcommand{\proj}[1]{\ket{#1}\!\bra{#1}}

\newcommand{\beq}{\begin{equation}}
\newcommand{\eeq}{\end{equation}}
\newcommand{\beqa}{\begin{eqnarray}}
\newcommand{\eeqa}{\end{eqnarray}}

\newtheorem{definition}{Definition}
\newtheorem{theorem}{Theorem}

\newtheorem{lemma}{Lemma}

\begin{document}
\begin{center}
{\LARGE\bf On the role of entanglement \\ in quantum computational
speed-up\\ }
\bigskip
{\normalsize Richard Jozsa$^\dagger$ and Noah Linden$^\S$}\\
\bigskip
{\small\it $^\dagger$Department of Computer Science,
University of Bristol,\\ Merchant Venturers Building, Bristol BS8 1UB U.K. \\

$^\S$Department of Mathematics, University of Bristol,\\
University Walk, Bristol BS8 1TW U.K.}
\\[4mm]
\date{today}
\end{center}

\begin{abstract}
For any quantum algorithm operating on pure states we prove that the
presence of multi-partite entanglement, with a number of parties that
increases unboundedly with input size, is necessary if the quantum
algorithm is to offer an exponential speed-up over classical
computation. Furthermore we prove that the algorithm can be
classically efficiently simulated to within a prescribed tolerance
$\eta$ even if a suitably small amount of global entanglement
(depending on $\eta$) is present. We explicitly identify the
occurrence of increasing multi-partite entanglement in Shor's
algorithm. Our results do not apply to quantum algorithms operating
on mixed states in general and we discuss the suggestion that an
exponential computational speed-up might be possible with mixed
states in the total absence of entanglement. Finally, despite the
essential role of entanglement for pure state algorithms, we argue
that it is nevertheless misleading to view entanglement as a key
resource for quantum computational power.

\end{abstract}

\section{Introduction}\label{intro}

Quantum computation is generally regarded as being more powerful than
classical computation. The evidence for this viewpoint begins with
Feynman's pioneering observation \cite{feynman} that the simulation
of a general quantum evolution on a classical computer appears to
require an exponential overhead in computational resources compared
to the physical resources needed for a direct physical implementation
of the quantum process itself. Subsequent work by Deutsch
\cite{deutsch}, Bernstein and Vazirani \cite{bv}, Simon \cite{simon},
Grover \cite{grover}, Shor \cite{shor} and others showed how quantum
evolution can be harnessed to carry out some useful computational
tasks more rapidly than by any known classical means. Perhaps the
most dramatic such result is Shor's quantum algorithm for integer
factorisation which succeeds in factoring an integer of $n$ digits in
a running time that grows less rapidly than $O(n^3 )$ whereas the
best known classical algorithm is exponentially slower (with running
time $O(\exp(n^{\frac{1}{3}} \log n ^\frac{2}{3}))$). Thus for some
computational tasks (such as factoring) quantum physics appears to
provide an exponential benefit but for other tasks (such as
satisfiability or other NP complete problems\cite{gj}) the quantum
benefits appear to be inherently more restricted\cite{bbbv} (giving
perhaps at most a polynomial speedup).

The concept of computational power provides a fundamentally new
language and set of tools for studying the relationship between
classical and quantum physics. Indeed it is of great interest to try
to characterise from this point of view, the essential non-classical
ingredients that give rise to the enhanced quantum computational
power. Also an understanding of the limitations of quantum
computational power (such as the apparent lack of an efficient
solution of an NP complete problem) may provide insights into the
strange architecture of the quantum formalism, perhaps even leading
to physical principles, formulated in terms of the concept of
computational complexity, that may guide the development of new
physical theories by severely restricting acceptable forms of
proposed new formalisms. Indeed arbitrarily created ``toy'' physical
theories (including most proposed non-linear generalisations of
quantum theory\cite{AL}) tend to engender immense computing power.
The apparently limited nature of {\em quantum} computational power
makes quantum theory rather atypical so this observation is probably
significant.

One fundamental non-classical feature of the quantum formalism is the
rule for how the state space $\cs_{AB}$ of a composite system $AB$ is
constructed from the state spaces $\cs_A$ and $\cs_B$ of the parts
$A$ and $B$ \cite{jozsa1}. In classical theory, $\cs_{AB}$ is the
cartesian product of $\cs_A$ and $\cs_B$ whereas in quantum theory it
is the tensor product (and the state spaces are linear spaces). This
essential distinction between cartesian and tensor products is
precisely the phenomenon of quantum entanglement {\em viz.} the
existence of pure states of a composite system that are not product
states of the parts.

In quantum theory, state spaces always admit the superposition
principle whereas in classical theory, state spaces generally do not
have a linear structure. But there do exist classical state spaces
with a natural linear structure (e.g. the space of states of an
elastic vibrating string with fixed endpoints) so the possibility of
superposition {\em in itself}, is not a uniquely quantum feature.

In \cite{jozsa1,ekertjozsa} it was argued that there is a
relationship between entanglement and the apparent ability of a
quantum process to perform a computational task with exponentially
reduced resources (compared to any classical process). We briefly
recall the two main points made in \cite{jozsa1,ekertjozsa}. The
first point concerns the physical resources needed to represent
superpositions in quantum versus classical theory. To represent a
superposition of $2^n$ levels classically, the levels must all
correspond to a physical property of a single system (with no
subsystems) as classical states of separate systems can never be
superposed. Hence we will need an exponentially high tower of levels
and the amount of the physical resource needed will grow {\em
exponentially} with $n$. In contrast, in quantum theory because of
entanglement, a general superposition of $2^n$ levels may be
represented in $n$ 2-level systems. Thus the amount of the physical
resource (that defines the levels) will grow only {\em linearly} with
$n$ (i.e. the number of 2-level systems). In the classical case one
may attempt to circumvent the need for an exponential tower by
considering a linear system with infinitely many levels that
accumulate below a finite upper bound. In this way we could represent
exponentially growing superpositions with only a constant cost in the
physical resource. But now the levels must crowd exponentially
closely together and we will need to build our instruments with
exponentially finer precision. This again will presumably require
exponentially increasing physical resources.

This first point may be expressed in more computational terms as
follows. Any positive integer $N$ may be represented in unary or
binary. The unary representation (being a string of 1's of length
$N$) is exponentially longer than the binary representation (having
length $\log N $). We may equivalently take the unary representation
as a string of $(N-1)$ 0's followed by a single 1, which we view as a
single mark at height $N$. Now consider physical implementations of
these representations of numbers. The unary representation
corresponds to the $N^{\rm th}$ level of a single system and hence we
can superpose unary representations of numbers in either classical or
quantum physics. For binary numbers we can exploit the compactness of
the representation by using $\log N$ 2-level systems. In that case we
can superpose these representations in quantum theory but not in
classical theory. In summary, physical representations of binary
number exist in both classical and quantum systems but only in the
quantum case can these representations be superposed. This is
precisely the phenomenon of quantum entanglement. If we wish to
perform computations on superpositions of numbers in a classical
setting then this is possible but we must use the exponentially more
costly unary representation i.e. the quantum formalism offers a far
greater potential power for computations in superposition.

The second point made in \cite{jozsa1} concerns the classical
computational cost of mimicking a typical step in a quantum
computation, which we epitomise as follows. Suppose that at some
stage of a quantum computation the state is an $n$-qubit (generally
entangled) state $\ket{\alpha} = \sum a_{i_1 \ldots i_n}
\ket{i_1}\ldots \ket{i_n}$ and suppose that the next step is the
application of a 1-qubit gate $U$ to the first qubit. The updated
state is then $\ket{\alpha'} = \sum a'_{i_1 i_2 \ldots i_n}
\ket{i_1}\ldots \ket{i_n}$ where \beq \label{calc} a'_{i_1 i_2 \ldots
i_n} = \sum_{j_1} U_{i_1 j_1} a_{j_1 i_2 \ldots i_n} \eeq and
$U_{ij}$ are the matrix elements of $U$. This is only one step of
quantum computation, on only $n$ qubits, but to classically compute
the updated state description via eq. (\ref{calc}) we would require
exponentially many arithmetic operations  and also the classical
description of the state itself is exponentially large (compared to
the compact $n$ qubit quantum physical representation). Now the point
is that these exponential gaps between classical and quantum
resources can be connected to the concept of entanglement as follows.
If the state $\ket{\alpha}$ were unentangled i.e. if $a_{i_1 \ldots
i_n}$ is given as a product $a_{i_1}b_{i_2} \ldots d_{i_n}$ then for
the classical representation, both the state description and the
computation of its update become polynomially sized, with $O(n)$
resources for the state description and a constant amount of
computation for the update, and these are now equivalent to the
corresponding quantum resources. This suggests that if entanglement
is absent in a quantum algorithm then the algorithm can be
classically simulated with an equivalent amount of classical
resources. Stated otherwise, if a quantum algorithm (on pure states)
is to offer an exponential speedup over classical algorithms, then
entanglement must appear in the states used by the quantum algorithm.

Our discussion above has been qualitative and we have glossed over
various significant issues.  Firstly, the terms in eq. (\ref{calc})
are complex numbers and from a computational point of view they have
potentially infinite sized descriptions.  Hence they will need to be
restricted to, or replaced by, suitably finitely describable numbers
(such as rationals) if we are to avoid a potentially prohibitively
large cost for the classical computation of individual arithmetic
operations.  The second issue concerns how much entanglement and what
type of entanglement is required if a quantum algorithm is to offer
exponential speed up over classical algorithms.  This is a main
concern of the present paper.  Let $p$ be any fixed positive integer.
A state of $n$ qubits will be called $p$-blocked if no subset of
$p+1$ qubits are entangled together.  Thus a $p$-blocked state may
involve a large amount of entanglement but this entanglement is
restricted, in its multi-partiteness, to sets of a most $p$ qubits.
Now suppose that a quantum algorithm has the property that the state
at each step is pure and $p$-blocked.  We note that the qubits in
each block can vary from step to step in the algorithm; indeed each
qubit can be entangled with every other qubit at some stage of the
computation.  Then we will show that again the algorithm can be
classically simulated with an amount of classical resources that is
equivalent to the quantum resources of the original algorithm (i.e.
the classical resources needed are polynomially, but not
exponentially, larger than the quantum resources).

We note that, in contrast, in communication tasks, entanglement
restricted to merely bi-partite entanglement suffices for exponential
benefits (for example an exponential reduction of communication
complexity \cite{comm}) but for ``standard" computational tasks our
results show that the availability of (even increasing amounts of)
only bi-partite entanglement cannot give an exponential speed-up
(despite the fact that 2-qubit {\em gates} suffice for universal
computation).  Also our results show that a distributed quantum
computer, which has any number of quantum processors but each with
bounded size and only classical communication between them, cannot
offer an exponential speed up over classical computation - if the
local processors have size up to $p$ qubits each then the state will
be $p$-blocked at every stage.

In section \ref{simul} we will prove our main result, that for any
quantum algorithm operating on pure states, the presence of
multi-partite entanglement, with a number of parties that increases
unboundedly with input size, is necessary if the quantum algorithm is
to offer an exponential speed-up over classical computation (theorem
\ref{pblthm}). Furthermore we will prove that the algorithm can be
classically efficiently simulated to within a prescribed tolerance
$\eta$ even if a suitably small amount of entanglement (depending on
$\eta$) is present (theorem \ref{entsim}). The presence or absence of
increasingly widespread multi-partite entanglement is not an obvious
feature of a given family of quantum states and in section
\ref{entshor} we will explicitly identify the presence of this
resource in Shor's algorithm.

The theorems of section \ref{simul} apply to quantum algorithms that
operate on {\em pure} states which are required to be $p$-blocked
(for some fixed $p$) at every stage. As such, the theorems actually
also apply to quantum algorithms on {\em mixed} states, where the
state is again required to be $p$-blocked at each stage (i.e. any
such algorithm can be classically efficiently simulated). Now, for
pure states, the $p$-blocked condition corresponds exactly to the
absence of entanglement (of more than $p$ qubits) but this is no
longer true for mixed states: a mixed state $\rho$ is $p$-blocked if
it is a product of mixtures (cf definition \ref{pblock}) whereas
$\rho$ is unentangled if it is separable i.e. a mixture of products
(of $p$-qubit states) and such $\rho$'s are not necessarily
$p$-blocked. Thus for mixed states, the $p$-blocked condition is
considerably stronger than the condition of absence of entanglement.
This leads to the following question: suppose that a quantum
algorithm operates on general mixed states and at every stage the
state is unentangled (in the sense of being separable). Can the
algorithm be classically simulated with an amount of classical
resources that is (polynomially) equivalent to the quantum resources
of the original algorithm? This fundamental question remains
unresolved. In section \ref{mixed} we will discuss some essential
differences between pure and mixed unentangled states which suggest a
negative answer (in contrast to the case of pure states) i.e. that an
exponential computational speed-up might plausibly be achievable in
quantum algorithms on mixed quantum states that have no entanglement.

According to our main results above  we can say that entanglement
is necessary in a quantum algorithm (on pure states) if the
algorithm is to offer an exponential speed-up over classical
computation. Does this mean that entanglement can be identified as
an essential resource for quantum computational power?  In section
\ref{what?} we will suggest (perhaps somewhat surprisingly) that
this is {\em not} a good conclusion to draw from our results!
Indeed our theorems are based entirely on the idea of
mathematically representing quantum states as {\em amplitudes} and
then classically computing the quantum evolution with respect to
{\em this} mathematical description. From our computational point
of view entanglement emerges as a significant property because
absence of entanglement implies a polynomially sized description
of the quantum process, {\em when the process is mathematically
expressed in  the amplitude description}. But suppose that instead
of the amplitude description we choose some other mathematical
description $\cd$ of quantum states (and gates). Indeed there is a
rich variety of possible alternative descriptions. Then there will
be an associated property $prop (\cd )$ of states which guarantees
that the $\cd $-description of the quantum computational process
grows only polynomially with the number of qubits if $prop (\cd )$
is absent (e.g. if $\cd$ is the amplitude description then $prop
(\cd )$ is just the notion of entanglement). Thus, just as for
entanglement, we can equally well claim that $prop (\cd )$ for
{\em any} $\cd$, is an essential resource for quantum
computational speed-up! Entanglement itself appears to have no
special status here. In summary, the point is that we can have
different mathematical formalisms for expressing quantum theory,
and although they are fully mathematically equivalent, they will
lead to quite different families of states (of increasingly many
qubits) that have a polynomially sized description with respect to
the chosen formalism. Hence we also get different notions of a
physical quality that guarantees a state will not be of the latter
form. Then {\em every} one of these qualities must be present in a
quantum algorithm if it is to offer an exponential speed-up over
classical algorithms.

\section{Preliminary definitions}\label{prelim}

We will need a precise definition of the notion of a quantum
computational process and a definition of what it means to
classically efficiently simulate such a process.

The term `state' will be used to mean a general (mixed) state. The
term $\poly (n)$ will refer to a function $f(n)$ whose growth is
bounded by a polynomial function in $n$ i.e. there exists a
polynomial $p(n)$ such that $f(n) \leq p(n)$ for all sufficiently
large $n$.

We adopt the gate array model of quantum computation as our working
definition. Let $\cg$ be a fixed finite set of 2-qubit gates.

{\definition \label{qcompproc} A {\bf quantum computational
process} (or quantum algorithm) with running time $T(n)$ comprises
the following description. For each fixed positive integer $n$
(input size) we have a sequence of triples \beq \label{triples}
\ca_n = \{ (U_{i_0}, a_0,b_0),\, (U_{i_1},a_1,b_1), \ldots ,
(U_{i_{T(n)}}, a_{T(n)},b_{T(n)}) \} \eeq where the $U_{i_j}$'s
are chosen from $\cg$ and $a_k$ and $b_k$ are positive integers.
More precisely, there is a classical algorithm running in $\poly
(T(n))$ time, which given $n$, will output the list $\ca_n$.

The quantum algorithm corresponding to the sequence $\{ \ca_n$:
$n=1,2, \ldots \}$ runs as follows. For each input $i_1 \ldots i_n$
of size $n$ we start with a row of qubits $\ket{i_1}\ket{i_2}\ldots
\ket{i_n} \ket{0}\ket{0} \ldots$ giving the input extended by zeroes.
We apply the sequence of $T(n)$ computational steps given in $\ca_n$,
where the $k^{\rm th}$ step is the application of the 2-qubit gate
$U_{i_k}$ to qubits $(a_k,b_k)$ in the row. After $T(n)$ steps we
measure the leftmost qubit in the computational basis and output the
result (0 or 1) i.e. we give a sample from the probability
distribution $\cp = \{ p_0,p_1 \}$ defined by the quantum measurement
on the final state.}

\noindent {\bf Remark} Later we will also discuss quantum
computational processes on {\em mixed} states. For such a process of
$T(n)$ steps we will require that the input state is a mixed state of
$n$ qubits with a $\poly(n)$ sized description (rather than just a
computational basis state of $n$ qubits as above). Also the
computational steps could be unitary or more generally, trace
preserving completely positive maps on two qubits. Equivalently we
could require the computational steps to be unitary transforms of six
qubits (i.e. having a 4-qubit ancilla space) whose locations are all
specified similar to the pairs of qubits in the above definition.

We will also need a notion of distance between states and between
probability distributions. For this purpose we will use the trace
norm\cite{bhatia}. For any operator $A$ (on a finite dimensional
state space), the trace norm $||A||$ is defined by  \[ ||A|| = \tr
\sqrt{AA^\dagger} = \sum_i \mu_i \] where $\mu_i$ are the singular
values of $A$. If $A$ is hermitian then $\mu_i$ are the absolute
values of the eigenvalues. If $\rho$ is a density matrix then
$||\rho||=1$.

The trace norm distance $||\rho-\sigma ||$ between states has an
especially useful property of being contractive under any trace
preserving quantum operation \cite{NC}. In particular if $\rho_{AB}$
and $\sigma_{AB}$ are bipartite states and $\rho_A , \sigma_A$ denote
the corresponding reduced states of $A$ then \[ ||\rho_A - \sigma_A
|| \leq ||\rho_{AB}-\sigma_{AB}||. \] Also if $\cp$ and $\cq$ are the
probability distributions for the outcomes of any quantum measurement
on two states $\rho$ and $\sigma$ respectively then \beq
\label{contprob} ||\cp -\cq || \leq ||\rho -\sigma || \eeq where
$||\cp -\cq || = \sum |p_i - q_i |$ is the trace norm distance
between the distributions viewed as diagonal states.

Finally we give our definition of the notion of classical efficient
simulation.

{\definition \label{simuldef}
 A quantum computation $\{ \ca_n : n=1,2, \ldots \}$ with output
 probability distribution $\cp$
 can be {\bf efficiently classically
simulated} if the following condition is satisfied:\\ Given only
classical means (i.e. a universal classical computer which is also
able to make probabilistic choices) and the description $\ca_n$ (i.e.
the classical poly-time algorithm for generating $\ca_n$ from $n$),
then for any $\eta >0$, we are able to sample once a distribution
$\cp'$ with $||\cp -\cp'|| \leq \eta$, using a classical
computational effort that grows polynomially with $n$ and $\log
(\frac{1}{\eta})$. }

In the above definition we should be more precise about the
computer's ability to make probabilistic choices. Just as
computational steps cannot be arbitrary, but must be chosen from a
fixed finite set of gates (giving a measure of computational effort
for any desired transformation) we need to have a measure of how much
computational effort is required to sample a given probability
distribution $\{ p_0 , p_1 \} $. We assume that the computer can only
toss a fair coin i.e. sample the probability distribution $\{
prob(0)=\frac{1}{2}, prob(1)=\frac{1}{2} \} $ and this counts as a
{\em single} computational step. Let $0\leq x\leq 1$ be any number
whose binary expression has at most $n$ binary digits: $x=0.i_1 i_2
\ldots i_n$. Then the computer can sample the distribution $\cp_x =
\{ x,1-x \}$ in $n$ steps as follows: toss the fair coin $n$ times
giving a sequence of results $j_1 \ldots j_n$. View $j_1 \ldots j_n$
as an $n$ digit binary number (and similarly $i_1 \ldots i_n = 2^n
x$). Then $prob(j_1 \ldots j_n <i_1 \ldots i_n ) =x$ so we get a
sampling of $\cp_x$ by comparing the random output $j_1 \ldots j_n$
to the given $i_1 \ldots i_n$. If $x$ has an infinite binary
expansion we can sample an $n$ digit approximation to $\cp_x$ by the
above method using $\poly(n)$ steps i.e. we can sample $\cp'$
having$||\cp' -\cp || \leq \eta$, with $\poly(\log 1/\eta )$
computational effort and we adopt the latter simulation rate as the
definition of efficient simulability for a general probability
distribution.

In many applications we do not need to consider arbitrarily small
$\eta$ as in definition \ref{simuldef} and a weaker simulation
requirement suffices. Suppose that the quantum computation is a BQP
algorithm for a decision problem. Thus for any input, the output
distribution $\cp = \{ p_0 ,p_1 \}$ has the property that the
probability of obtaining a correct answer is $>\frac{2}{3}$. In that
case the decision problem will have a classical efficient (BPP)
algorithm if we can efficiently classically simulate a distribution
$\cp' = \{ p_0' , p_1' \}$ with $|| \cp - \cp' || \leq \eta_0$ where
$\eta_0  <\frac{1}{6}$ is a constant, so that the probability of a
correct answer with $\cp'$ is still bounded away from $\frac{1}{2}$.
Indeed we will see (theorem \ref{entsim} below) that for such a
finite tolerance classical simulation to be ruled out, not only must
the quantum algorithm exhibit multi-partite entanglement in its
states but furthermore the amount of this entanglement must be
suitably large (with a lower bound depending on $\eta_0$ and the
running time of the algorithm).

\section{Simulation by classical computation}\label{simul}

One method for classically simulating a quantum computation  is to
directly compute the state at each step from the sequence of unitary
operations prescribed in the quantum algorithm. We will investigate
the implications of this simulation for the power of quantum
computing compared to classical computing, especially noting
conditions which guarantee that this simulation is efficient.

Let $\ket{\alpha_j}$ be the state after $j$ steps of computation,
which we may assume is a general state of at most $2j$ qubits (by
neglecting unused qubits from the initial row). In the
computational basis we have: \beq \label{psij} \ket{\alpha_j} =
\sum a_{i_1 i_2 \ldots i_{2j}} \ket{i_1}\ket{i_2} \ldots
\ket{i_{2j}}. \eeq Then $\ket{\alpha_{j+1}}$ is obtained by
applying the 2-qubit gate $U_{i_{j}}$ to qubits $a_{j}$ and
$b_{j}$. For clarity let us assume that these are the first two
qubits i.e. $a_{j}=1$ and $b_{j}=2$ (all other possible cases are
similar). The amplitudes $\tilde{a}$ of the updated state are
calculated as: \beq \label{update} \tilde{a}_{i_1 i_2 i_3 \ldots
i_{2j}} = \sum_{j_1,j_2 = 0,1} M^{j_1j_2}_{i_1i_2}a_{j_1 j_2 i_3
\ldots i_{2j}} \eeq where $ M^{j_1j_2}_{i_1i_2}$ is the 4 by 4
matrix of $U_{i_{j}}$.

This classical computation may fail to be efficient for two
reasons. Firstly there are exponentially many amplitudes that need
to be computed and the matrix multiplication of $M$ in eq.
(\ref{update}) needs to be carried out exponentially many times.
This inefficiency is intimately related to the fact that the
states $\ket{\alpha_j}$ are generally entangled and the
implications of this obstacle to efficient simulation will be
elaborated below.

The second possible difficulty with the computation in eq.
(\ref{update}) arises from that fact that the matrix entries of
$M$ are generally continuous parameters (real or complex numbers)
so that even the individual arithmetic operations (additions and
multiplications) involved might be prohibitively costly. This
second issue will be circumvented by considering rational
approximations to gates.

{\definition \label{rat} A quantum gate is {\bf rational} if its
matrix elements (in the computational basis) have rational numbers as
real and imaginary parts. }

The main property of rational numbers that we will need is the
following.

{\lemma \label{ratlemma} Let $\cd = \{ r_1, \ldots ,r_L \}$ be a
finite set of rational numbers whose numerators and denominators have
at most $m$ digits. For any $j$ let $x$ be an arithmetic expression
constructed from the elements of $\cd$ using at most $j$ additions
and multiplications. Then $x$ can be computed exactly (as a rational
number) using a number of steps of computation that grows
polynomially with $j$ and $m$.}
\\ {\bf Proof\, } This is an easy consequence of the polynomial
(in the number of digits) computability of integer arithmetic and of
the fact that the number of digits of the numerator and denominator
of an arithmetic expression in the rationals in $\cd$ grows at most
linearly with the number of operations.$\,\, QED$.

Apart from rational gates, many other possible classes of gates,
having the essential polynomial property of lemma \ref{ratlemma},
would suffice for our purposes. For example we could allow the matrix
elements to be members of a finite algebraic extension of the
rationals. This would allow numbers such as $\frac{1}{\sqrt{2}}$ and
$\cos \pi/8$ which appear as matrix elements of commonly used
universal sets of gates \cite{NC}.

Let us now return to the relation of entanglement to the
efficiency of the classical computation in eq. (\ref{update}).

{\definition \label{pblock} Let $\rho$ be a state of $m$ qubits where
the qubits are labelled by $B= \{ 1,2, \ldots ,m \}$. $\rho$ is {\bf
$p$-blocked} if $B$ can be partitioned into subsets of size at most
$p$:
\[ B= B_1 \cup B_2 \cup \ldots \cup B_K \hspace{1cm} |B_i | \leq p
\] and $\rho$ is a product state for this partition:
\[ \rho = \rho_1 \otimes \rho_2 \otimes \ldots \otimes \rho_K \]
where $\rho_i$ is a state of the qubits in $B_i$ only.}

Note that a pure state is $p$-blocked if and only if no $p+1$
qubits are all entangled together. But a mixed state $\rho$ can be
unentangled in the sense of being separable, without being
$p$-blocked i.e. the $p$-blocked condition (``product of
mixtures'') is stronger than the separability condition (``mixture
of products'').

Our next result transparently shows the necessity of multi-partite
entanglement for exponential quantum computational speed-up in a
restricted situation where there are no additional complications
arising from the description of the gates (as they are assumed to
be rational).

{\lemma \label{ratpbl} Let $\cg$ be a finite set of rational 2-qubit
gates and $p$ a fixed positive integer. Suppose that $\{ \ca_n :
n=1,2,\ldots \}$, using gates from $\cg$, is a polynomial time
quantum computation with the property that at each stage $j=1,\ldots
,\poly (n)$ the state $\ket{\alpha_j}$ is $p$-blocked. Then the final
probability distribution $\cp$ can be classically exactly computed
with $\poly (n)$ computational steps so the quantum computation can
be classically efficiently simulated.}

\noindent {\bf Proof of lemma \ref{ratpbl}\,} Any $p$-blocked pure
state $\ket{\psi}$ of $m$ qubits may be fully described with the
following data: \\ (a) (Block locations) A list of $m$ integers
$(b_1, \ldots ,b_m)$ where $1\leq b_i \leq m$. $b_i$ gives the number
of the block to which the $i^{\rm th}$ qubit belongs. For example the
list $(3,5,4,3,3, \ldots )$ denotes that qubit 1 is in block 3, qubit
2 is in block 5 and so on. Note that the number of blocks can grow at
most linearly with $m$.\\ (b) (Block states) For each block we give
its state by listing the amplitudes in the computational basis. This
requires at most $2^{p+1}$ real numbers since each block has size at
most $p$ qubits.

Note that for fixed $p$ and increasing $m$ the total size of the
description grows only polynomially with $m$ assuming that the real
numbers in (b) can each be described with $\poly (m)$ bits of memory.
This is in contrast to the exponentially growing number of amplitudes
needed to describe a general entangled state.

Now to classically simulate the $p$-blocked algorithm we simply
update the ((a),(b)) description of the state at each step. Note that
the location of the blocks (i.e. (a)) will generally change as well
as the states of the blocks themselves (i.e. (b)). The $j^{\rm th}$
computational step is given by the action of a 2-qubit gate $U_{i_j}$
on a $p$-blocked state and we distinguish two cases:\\ Case 1: the
gate acts on two qubits which are already in the same block. Thus (a)
remains unchanged and the state of the chosen block is updated in (b)
by applying the unitary matrix of size at most $2^p \times 2^p$,
requiring a constant number of arithmetic operations (which does not
grow with $j$, the counter describing the step of the quantum
algorithm that is being simulated).\\ Case 2: the 2-qubit gate
straddles two existing blocks $B_1$ and $B_2$ of sizes $p_1 \leq p$
and $p_2 \leq p$ respectively. We again update the state of all $p_1
+ p_2$ qubits by applying a unitary matrix of size at most $2^{2p}
\times 2^{2p}$ (requiring a constant number of operations that does
not grow with $j$). If $p_1 + p_2 \leq p$ we also update (a) by
amalgamating the two block labels into a single label, to complete
the step. If $p_1 + p_2 >p$ we need to identify a new block structure
(with blocks of size $\leq p$) within the qubits of $B_1$ and $B_2$.
One method is to compute the reduced state $\rho_X$ of every subset
$X\subseteq B_1 \cup B_2$ having size $\leq p$, and compare the
global state of $B_1 \cup B_2$ with $\rho_{X_1} \otimes \ldots
\otimes \rho_{X_K}$ for each partition $X_1 ,\ldots ,X_K$ of $B_1
\cup B_2$ (looking for an equality of states). This calculation again
clearly needs a bounded number of arithmetic operations (independent
of $j$). Finally we update (a) and (b) with the newly found blocks
and their corresponding states.

This gives a classical simulation of the quantum algorithm using a
number of rational arithmetic operations that grows linearly with
$j$. Finally lemma \ref{ratlemma} guarantees that the calculation can
be done with $\poly (j)$ elementary computational steps, giving an
efficient classical computation of the final state of the quantum
algorithm. Finally we identify the block containing the leftmost
qubit, compute the probability distribution $\cp$ and sample it once
with a corresponding classical probabilistic choice, completing the
efficient classical simulation of the quantum algorithm. $\, QED$.

\noindent {\bf Remark} Although lemma \ref{ratpbl} has been stated
for quantum algorithms with {\em pure} states it is readily
generalised to the case of mixed states: suppose that a quantum
algorithm (with rational gates) has a $p$-blocked {\em mixed} state
at each stage. Then it can be classically efficiently simulated.
Indeed the above proof does not require the block states to be pure.
In a similar way, the theorems below also easily generalise to
$p$-blocked mixed state processes (although for clarity we give the
statements and proofs only for the case of pure states).

Lemma \ref{ratpbl} depends on two essential ingredients: (a)
$p$-blockedness implying a polynomial number of parameters for state
descriptions and (b) rationality of gates, guaranteeing that the
classical arithmetic operations can be efficiently carried out. Our
next result shows that the condition (b) can be lifted.

{\theorem \label{pblthm} Let $\cg$ be a finite set of 2-qubit gates
and $p$ a fixed positive integer. Suppose that $\{ \ca_n :
n=1,2,\ldots \}$, using gates from $\cg$, is a polynomial time
quantum computation with the property that at each stage $j=1,\ldots
,\poly (n)$ the state $\ket{\alpha_j}$ is $p$-blocked. Then the
quantum computation can be classically efficiently simulated.}

Before going on to consider the proof of theorem \ref{pblthm} we make
some remarks on the significance of this result. Theorem \ref{pblthm}
shows that {\em multi-partite} entanglement with {\em unboundedly
many} qubits entangled together, is a necessary feature of any
quantum algorithm (operating on pure states) if the algorithm is to
exhibit an exponential speed-up over classical computation. Indeed
absence of increasing numbers of entangled qubits corresponds to a
fixed value of $p$. In contrast, in communication tasks, entanglement
restricted to merely bi-partite entanglement suffices for exponential
benefits (for example an exponential reduction of communication
complexity \cite{comm}) but for ``standard'' computational tasks the
availability of (even increasing amounts of) only bi-partite
entanglement cannot give an exponential speed-up (in contrast to the
fact that 2-qubit {\em gates} suffice for universal computation)
--- the number of qubits entangled together must grow as an
unbounded function of input size. Indeed even if every pair of qubits
become entangled at some stage of the computation and there is no
higher order entanglement, the quantum algorithm will still have an
efficient classical simulation. This shows that the role of
entanglement in computation is essentially different from its role in
communication. Theorem \ref{pblthm} also implies that distributed
quantum computing (on pure states), which allows any number of local
quantum processors, but of {\em bounded size} and classical
communication between them, cannot offer an exponential speed-up over
classical computation -- if the local processors have size up to $p$
qubits each then the state will be $p$-blocked at every stage.

Our approach to proving theorem \ref{pblthm} will be to replace
general gates by rational approximations with increasingly high
precision. Recall \cite{barenco} that there exists a finite universal
set of rational 2-qubit gates. Hence any quantum computation $QC_1$
with gate set $\cg_1 = \{ U_i \}$ can be efficiently approximated by
a quantum computation $QC_2 (\epsilon )$ having rational gates $\cg_2
= \{ \tilde{U}_i \}$ where
 $|| U_i - \tilde{U}_i || <\epsilon$ (for any chosen
$\epsilon >0$). As far as a quantum {\em physical} implementation is
concerned, $QC_2$ behaves very similarly to $QC_1$ in its action on
states \cite{vazirani}. But for our classical {\em computational}
simulations there can be a dramatic difference: if $QC_1$ is a
$p$-blocked computation (and hence efficiently simulable by the
theorem) then the approximation $QC_2$ will generally {\em not} be
$p$-blocked so we cannot invoke the lemma to claim that $QC_2$
(although near to $QC_1$ and having rational gates) is efficiently
simulable. Indeed the fact that $p$-blocked states of $m$ qubits have
a $\poly(m)$ sized exact description (which is crucial in the proof
of lemma \ref{ratpbl}) is immediately lost under arbitrarily small
perturbations to general states, which require an exponentially large
description (of exponentially many independent amplitudes). Hence the
theorem is not a straightforward consequence of the lemma via
efficient rational approximation of the gates, and as will be seen,
its proof will require substantial further ingredients. Our strategy
will be to further modify $QC_2$ to a nearby process $QC_2'$ whose
states are $p$-blocked transmogrifications of the states of $QC_2$.
This leads us to consider the simulation of algorithms whose states
remain suitably close to $p$-blocked states, so they may have a small
amount of entanglement between the blocks. Consequently we will
develop suitable approximations with polynomially sized descriptions,
to states that are not $p$-blocked, but are suitably near to
$p$-blocked states. This is the content of theorems \ref{entsim} and
\ref{entsimrat} below (and theorem \ref{pblthm} will appear as a
special case).

Although we are using the gate array model of quantum computation for
our arguments, our results will apply to other models as well.
Suppose we have any model of quantum computation with the following
properties: \\ (a) At each stage of the computation we have a pure
state of a system comprising subsystems of a bounded size; \\ (b) The
update of the state is effected by a unitary transform or by a
measurement, each on a bounded number of subsystems. \\ In that case
our proofs readily generalise to show that if the states are
$p$-blocked at every stage then the computational process will have
an efficient classical simulation i.e. that multi-partite
entanglement of unboundedly many subsystems must be present for an
exponential computational speed-up. The above criteria for a
computational model are satisfied by the quantum turing machine model
\cite{bv} as well as some recently proposed models that focus on
measurement operations \cite{nielsen,briegel}.

As mentioned above we will prove theorem \ref{pblthm} in a more
general form motivated as follows. Let $\ket{\alpha_j}$ be the pure
state at step $j$ in a quantum algorithm. Write $\alpha_j$ as an
abbreviation for $\proj{\alpha_j}$. Theorem \ref{pblthm} requires
that $\alpha_j$ be exactly $p$-blocked for each $j$. Suppose that
$\alpha_j$ is not $p$-blocked but is made up of blocks of size at
most $p$ with a ``small'' amount of entanglement between the blocks.
We formalise this by requiring that for each $j$ there is a
$p$-blocked (possibly mixed) state $\beta_j$ close to $\alpha_j$:
 \beq \label{areclose} || \alpha_{j} - \beta_{j} ||
\leq \epsilon. \eeq

Theorem \ref{pblthm} states that if $\epsilon =0$ then a classical
efficient simulation exists. Our question now is: how large can
$\epsilon$ be in eq. (\ref{areclose}) so that the quantum
algorithm can still be efficiently classically simulated? i.e. we
wish to study the stability of efficient simulability under small
perturbations of the $p$-blockedness condition. The following
theorem says that if we accept an error tolerance $\eta$ in the
output probability distribution, then the quantum algorithm will
have an efficient classical simulation (to within the tolerance)
for nonzero $\epsilon$ exponentially small in the running time
($\epsilon \sim \eta c^T$ for $0<c<1$).

{\theorem \label{entsim} Let $\cg$ be a finite set of 2-qubit gates
and $p$ a fixed positive integer. Suppose that $\{ \ca_n : \, n= 1,2,
\ldots \} $ using gates from $\cg$, is a polynomial time quantum
computation with running time $T= \poly (n)$. Let $\cp$ be the output
probability distribution of the algorithm. For $j= 0, \ldots ,T$ let
$\ket{\alpha_j}$ denote the state at stage $j$ and write
$\proj{\alpha_j} = \alpha_j$. \\ Suppose that the states $\alpha_j$
are not exactly $p$-blocked but there exists a sequence of
$p$-blocked states $\beta_j$ (generally depending on $n$ too) such
that \beq \label{near} ||\alpha_{j} -\beta_{j} || \leq \epsilon. \eeq
(The identities of the states $\beta_j$ are not assumed to be known).
 \\ Then for any
$\eta >0$, if $\epsilon \leq \frac{\eta}{4(2p+4)^T}$, we can
classically sample a distribution $\cp'$  with $||\cp - \cp' || \leq
\eta$ using $\poly (T,\log 1/\eta )$ classical computational steps.}

Theorem \ref{pblthm} is an immediate consequence of theorem
\ref{entsim} (by taking $\epsilon =0$ in the latter theorem). To
prove theorem \ref{entsim} we first consider an analogue (theorem
\ref{entsimrat} below) with rational gates that may vary with $n$.
Then our strategy for theorem \ref{entsim} will be to replace
gates by rational approximations, but these approximations will
need to become suitably more accurate as $n$ increases. Theorem
\ref{entsimrat} may also be viewed as a generalisation of lemma
\ref{ratpbl}, allowing a small amount of inter-block entanglement.
If the states $\alpha_j$ in lemma \ref{ratpbl} are only suitably
close to being $p$-blocked (rather than being exactly $p$-blocked)
then a probability distribution $\cp'$ suitably close to $\cp$ can
still be efficiently calculated.

{\theorem \label{entsimrat}
 Suppose that $\{ \ca_n : \, n=
1,2, \ldots \} $  is a polynomial time quantum computation with
running time $T= \poly (n)$ and using only rational gates. Let $m$
(which may grow with $n$) be the largest number of digits of the
denominators and numerators of the rational gates used in $\{ \ca_k :
1\leq k\leq n \}$.
 Let
$\cp$ be the output probability distribution of the algorithm. For
 $j= 0, \ldots ,T$ let $\ket{\alpha_j}$ denote the state at stage
$j$ and write $\proj{\alpha_j} = \alpha_j$. \\ Suppose that the
states  $\alpha_j$ are not exactly $p$-blocked but there exists a
sequence of $p$-blocked states $\beta_j$ (generally depending on
$n$ too) such that  \beq \label{nearprop} ||\alpha_{j} -\beta_{j}
|| \leq \epsilon. \eeq (The identities of the states $\beta_j$ are
not assumed to be known).
 \\ Then for any
$\eta >0$, if $\epsilon \leq \frac{\eta}{ (2p+4)^T}$, we can
classically sample a distribution $\cp'$  with $||\cp - \cp' ||
=\eta$ using $\poly (T,m )$ classical computational steps.}

\noindent {\bf Proof of theorem \ref{entsimrat}}\, We have a
computational process
\[ \ket{\alpha_0} = \ket{i_1}\ldots \ket{i_n} \ket{0}\ket{0} \ldots
\hspace{1cm} \ket{\alpha_{j+1}}= U_j \ket{\alpha_j} \hspace{1cm} j=0,
\ldots ,T \] where each $U_j$ is a rational 2-qubit gate. Using the
existence of the sequence $\beta_j$ we will show that there is a
sequence of states $\rho_j$ and numbers $e_j$ with the following
properties: \\ (a) $||\rho_j -\alpha_j || \leq e_j$. \\ (b) $\rho_j$
is $p$-blocked. \\ (c) $\rho_j$ can be classically computed in $\poly
(j,m)$ steps. \\ (d) $e_0 =0$ and $e_{j+1} \leq (2p+3) (e_j +\epsilon
)$ .
\\ (Note that the $p$-blocked states $\beta_j$ are not assumed to be
known and they will not generally satisfy the crucial condition (c)).

Assuming all the above is given, we see from (d) that $e_1 \leq
(2p+3) \epsilon$ and $e_{j+1} \leq (2p+4)e_j$ so \[ e_j \leq \epsilon
(2p+4)^j. \] Hence, given $\eta$, if we choose any $\epsilon \leq
\eta (2p+4)^{-T}$ we will have
\[ ||\rho_T - \alpha_T || \leq e_T \leq \eta. \] Consequently by
eq. (\ref{contprob}): \[ || \cp - \cp' || \leq \eta \] where
$\cp'$ is the probability distribution arising from a quantum
measurement on $\rho_T$. Hence by (c), $\cp'$ can be efficiently
classically sampled, as required.

The states $\rho_j$ are calculated sequentially as follows. Set
$\rho_0 = \alpha_0$ so $||\rho_0 - \alpha_0 || \leq e_0 =0$. Suppose
we have generated a $p$-blocked state $\rho_j$ for the $j^{\rm th}$
step with \[ ||\rho_j -\alpha_j || \leq e_j. \]  We construct
$\rho_{j+1}$ as follows. Let $\tau = U_j \rho_j U_j^\dagger$, with
the 2-qubit rational gate $U_j$ acting on qubits from $C= C_1 \cup
C_2$ where $C_1$ and $C_2$ are blocks in $\rho_j$ (so $|C| \leq 2p$).
$\tau$ might not be a $p$-blocked state if $|C|
>p$ but outside of $C$ $\tau$ remains $p$-blocked because $\rho_j$
was $p$-blocked (and $\tau$ and $\rho_j$ agree outside $C$). Since
$\alpha_{j+1}=U_j \alpha_j U_j^\dagger$ too, we have
$||\alpha_{j+1}-\tau || = ||\alpha_j -\rho_j || \leq e_j$ so \beq
\label{tautobeta} ||\tau -\beta_{j+1}|| \leq e_j +\epsilon. \eeq For
any subset $X$ of qubit positions let $\tau_X$ (respectively
$\rho_{Xj}$ and $\beta_{X(j+1)}$) denote the reduced state of $X$ in
$\tau$ (respectively $\rho_j$ and $\beta_{j+1}$). We aim to decompose
$C$ into $K\leq 2p$ blocks $E_i$ of size at most $p$ so that
$||\tau_C - \tau_{E_1} \otimes \ldots \otimes \tau_{E_K}||$ remains
suitably small.

Now $\beta_{j+1}$ is $p$-blocked so the reduced state
$\beta_{C(j+1)}$ is $p$-blocked too. Hence
$\beta_{C(j+1)}=\beta_{D_1(j+1)} \otimes \ldots \otimes \beta_{D_K
(j+1)}$ where $C=D_1 \cup \ldots \cup D_K$, $|D_i| \leq p$ and $K\leq
2p$.  From eq. (\ref{tautobeta}) we have $||\beta_{C(j+1)} -\tau_C ||
\leq e_j+\epsilon$ and \[ ||\beta_{D_i (j+1)} - \tau_{D_i}|| \leq e_j
+\epsilon \hspace{1cm} \mbox{for $i=1, \ldots ,K$}. \] Thus (cf the
hybrid argument of \cite{vazirani}) \[ || \beta_{D_1 (j+1)}\otimes
\ldots \beta_{D_K (j+1)}- \tau_{D_1} \otimes \ldots \otimes
\tau_{D_K}|| \leq K(e_j +\epsilon) \leq 2p(e_j +\epsilon ) \] and
using the triangle inequality \beq \label{close} \begin{array}{rcl}
||\tau_C -\tau_{D_1} \otimes \ldots \otimes \tau_{D_K}|| & \leq &
||\tau_C - \beta_{C(j+1)}|| + ||\beta_{C(j+1)}- \beta_{D_1
(j+1)}\otimes \ldots \beta_{D_K (j+1)}|| \\ & & + || \beta_{D_1
(j+1)}\otimes \ldots \beta_{D_K (j+1)} - \tau_{D_1} \otimes \ldots
\otimes \tau_{D_K}|| \\ & \leq & (e_j + \epsilon) + 0 + 2p(e_j
+\epsilon) \\ & = & (2p+1)(e_j+\epsilon). \end{array} \eeq Hence
there {\em exists} a partition of $C$ such that the corresponding
product of reduced states of $\tau$ approximates $\tau_C$ to within
$(2p+1)(e_j+\epsilon)$. Thus we compute $||\tau_C -\tau_{E_1} \otimes
\ldots \otimes \tau_{E_L}||$ for all partitions $C= E_1 \cup \ldots
\cup E_L$ of $C$ and choose one satisfying eq. (\ref{close}). Finally
to get $\rho_{j+1}$ we update $\tau_C$ by the chosen $ \tau_{E_1}
\otimes \ldots \otimes \tau_{E_L}$, giving a $p$-blocked state with
\[ \begin{array}{rcl} ||\rho_{j+1}-\alpha_{j+1} ||
& \leq &  ||\rho_{j+1}-\tau || + ||\tau -\beta_{j+1}|| +||
\beta_{j+1}-\alpha_{j+1}||  \\  & \leq & (2p+1)(e_j+\epsilon)+ (e_j
+\epsilon) +\epsilon \\ & \leq &  (2p+3) (e_j+\epsilon)
\end{array} \]
i.e. $e_{j+1} \leq (2p+3)(e_j+\epsilon)$ as required. The entire
calculation in updating $\rho_j$ to $\rho_{j+1}$ is
 carried out within a block $C$ of size at most $2p$ using only
 rational arithmetic operations and the number of operations does not
 grow with $j$. Hence by lemma \ref{ratlemma}, $\rho_T$ is
 calculated using a number of computational steps
 that grows polynomially with
 $T$ and $m$.\, $QED.$

\noindent {\bf Proof of theorem \ref{entsim}} \, Let $\tilde{U}_j$ be
rational gates with \beq \label{xi} ||\tilde{U}_j-U_j || \leq \xi
\eeq (with $\xi$ to be chosen later to match $\eta$ and $T$) and the
rational matrix elements of $\tilde{U}_j$ have numerators and
denominators with $O(\log \frac{1}{\xi} )$ digits, which is always
possible by eq. (\ref{xi}) (and the use of a universal rational
2-qubit gate \cite{barenco}). Consider the process \beq \label{tilde}
\tilde{\alpha}_0 =\alpha_0 \hspace{1cm} \tilde{\alpha}_{j+1} =
\tilde{U}_j \tilde{\alpha}_j \tilde{U}_j^\dagger \hspace{1cm}
j=0,1,\ldots ,T-1 \eeq which we will compare to the process with
$\alpha_{j+1}=U_j \alpha_j U_j^\dagger$.

If $||\tilde{\alpha}_j - \alpha_j || =e_j$, writing $A_j =
\tilde{U}_j - U_j$ we have $||A_j || \leq \xi$ and \[
\begin{array}{rcl}
||\tilde{\alpha}_{j+1}-\alpha_{j+1}|| & =  & ||
(U_j+A_j)\tilde{\alpha}_j (U_j^\dagger + A_j^\dagger ) -U_j
\alpha_j U_j^\dagger ||\\ & = & || U_j (\tilde{\alpha}_j -\alpha_j
)U_j^\dagger +A_j \tilde{\alpha}_j U_j^\dagger + U_j
\tilde{\alpha}_j A_j^\dagger + A_j \tilde{\alpha}_j A_j^\dagger ||
\\ &  \leq & e_j +\xi +\xi+\xi^2 \end{array} \]
where we have used the properties \cite{bhatia} $||UAV||=||A||$ and
$||AB||\leq ||A|| \, ||B||$ for any unitary $U,V$ and arbitrary
$A,B$. Thus $e_{j+1} = ||\tilde{\alpha}_{j+1}-\alpha_{j+1} || \leq
e_j +3\xi$. Hence \beq \label{tildetoalpha} ||\tilde{\alpha}_j
-\alpha_j || \leq 3j\xi \hspace{1cm} j=0,\ldots ,T-1. \eeq If
$\tilde{\cp}$ (respectively $\cp$) is the output distribution of the
final measurement of the quantum algorithm performed on
$\tilde{\alpha}_T$ (respectively $\alpha_T$) we get \beq
\label{endprob} ||\tilde{\cp} -\cp || \leq 3T\xi. \eeq Now
$\tilde{\alpha}_j$ are generated with rational gates but they are not
necessarily $p$-blocked. However they lie close to the $p$-blocked
states $\beta_j$. From $||\alpha_j -\beta_j || \leq \epsilon$ and eq.
(\ref{tildetoalpha}) we get \[ ||\tilde{\alpha}_j -\beta_j || \leq
3j\xi+\epsilon \leq 3T\xi+\epsilon \] and we can apply theorem
\ref{entsimrat} to claim the following: given any $\eta >0$ (and
writing $c= \frac{1}{(2p+4)}$), if $\xi$ and $\epsilon$ are chosen so
that \beq \label{xieps} 3T\xi +\epsilon \leq \eta c^T \eeq then we
can sample a distribution $\cp'$ with $||\tilde{\cp} -\cp'||\leq
\eta$ using $\poly (T, \log \frac{1}{\xi} )$ classical computational
steps. From eq. (\ref{endprob}) we have
\[ ||\cp -\cp' || \leq 3T\xi +\eta \leq \eta c^T +\eta \leq 2\eta.
\] Now we can satisfy eq. (\ref{xieps}) by fixing $\xi =
\frac{1}{6T} \eta c^T$ and letting $\epsilon < \frac{1}{2}\eta
c^T$ so $\log \frac{1}{\xi} = \log \frac{1}{\eta} +\poly (T)$ and
$\poly (T,\log \frac{1}{\xi} )= \poly (T, \log \frac{1}{\eta} )$.
Finally replacing $\eta$ by $\eta/2$ in the above, we see that for
any $\eta >0$, if $\epsilon <\frac{1}{4}\eta c^T$ then we can
sample $\cp'$ with $||\cp - \cp' ||\leq \eta$ using $\poly (T,\log
\frac{1}{\eta} )$ classical computational steps, as required.\,
$QED.$

\section{Multi-partite entanglement in Shor's
algorithm}\label{entshor}

Shor's algorithm \cite{shor} is generally believed to exhibit an
exponential speed-up over any classical factoring algorithm. Thus, in
the light of the arguments given above, one would expect that there
is entanglement of an unbounded number of particles in Shor's
algorithm. This is indeed the case as we now show.  Of course, this
does not show that there is no classical polynomial algorithm for
factoring; however if it had been the case that only a bounded number
of particles had been entangled for any value of the input size, then
the results in the previous sections would have furnished a classical
polynomial-time algorithm for factoring.

To see that an unbounded number of particles becomes entangled in
Shor's algorithm, it suffices to show that this happens at some point
in the algorithm. The key construction of the algorithm is a method
for determining the period $r$ of the function
\begin{equation}  f(x)=a^x\,\,\, \mbox{mod
$N$} \label{exponential-function}
\end{equation} where $N$ is the input number to be factorised and
$a<N$ has been chosen at random. Following a standard description of
the algorithm (e.g. as given in \cite{NC}) we see that at an
appropriate stage we will have a periodic state of about $2\log N$
qubits, of the form
\begin{eqnarray}
\sum_k |x_0+kr\rangle,
\end{eqnarray} where $0<x_0 <r$ is unknown and has been chosen at
random (and we have omitted the normalisation factor).

We will now show that, with high probability, these ``arithmetic
progression" states have unbounded numbers of particles entangled as
$N$ increases. In order to see how the argument proceeds, consider
first a case in which an arithmetic progression state {\em is}
blocked.  This case is the state (with $r=3$)
\begin{eqnarray}
|AP_1\rangle =\sum_{k=0}^3 |3+3kr\rangle = |3\rangle + |6\rangle
+|9\rangle +|12\rangle.
\end{eqnarray}
Expressing each state in binary we have
\begin{eqnarray}
|AP_1\rangle = |0_3 0_2 1_1 1_0\rangle + |0_3 1_2 1_1 0_0\rangle
+|1_3 0_2 0_1 1_0\rangle +|1_3 1_2 0_1 0_0\rangle.
\end{eqnarray}
We have labelled the qubits with a subscript indicating the power of
two to which it refers.  We now re-order the qubits to make it easy
to see that $|AP_1\rangle$ is blocked (into two blocks each
containing two qubits):
\begin{eqnarray}
|AP_1\rangle &=& |0_3  1_1 0_2 1_0\rangle + |0_3  1_1 1_2 0_0\rangle
+|1_3  0_1 0_2 1_0\rangle +|1_3  0_1 1_2 0_0\rangle\nonumber\\ &=&
\left(|0_3  1_1 \rangle + |1_3  0_1 \rangle\right) \otimes\left(| 0_2
1_0\rangle +| 1_2 0_0\rangle\right).
\end{eqnarray}

The demonstration that this state is blocked proceeded by identifying
which qubits are in which block.  Let us now consider a general
arithmetic progression.  Imagine that it can be $p$-blocked.  Thus we
can rearrange the ordering of the qubits so that the state may be
written
\begin{eqnarray}
\left( |a_1\rangle + |a_2\rangle + \ldots |a_{p_1}\rangle\right)
\otimes  \left( |b_1\rangle + |a_2\rangle + \ldots
|b_{p_2}\rangle\right) \otimes \ldots \otimes \left( |z_1\rangle +
|z_2\rangle + \ldots |z_{p_k}\rangle\right);
\end{eqnarray}
For example if $\ket{a_2}$ is $\ket{1_3 0_1}$ then $a_2 =8$. Each
term $a_i$ in the first bracket will a binary expression where
non-zero digits all lie in a given subset of positions (of size at
most $p$) i.e. the block corresponding to the first bracket.
Different round brackets correspond to disjoint such blocks of digit
positions.

We will arrange the terms in each round bracket in increasing
order of the binary string labelling the state. The full state in
the progression with lowest binary string  is $|a_1\rangle |
b_1\rangle \ldots | z_1\rangle$; i.e. the binary string is $a_1+
b_1 \ldots + z_1$. The next lowest term in the progression is $a_1
+ b_1 \ldots + z_1 + r$. Thus one of the brackets must have the
property that the difference between the  two smallest binary
strings is $r$. Let us say that this bracket is the one containing
the $|a_i\rangle$ (i.e. $a_2 -a_1 =r$). Remember that all the
binary numbers in the superposition in that bracket must be
expressible using a given set of up to $p$ bits (where $p$ does
not increase with $N$). Now as $N$ increases, so will the typical
values of $r$. A typical $r$ will contain the pair ``10" at two
adjacent places in its binary representation many times (typically
one quarter of the time). For large enough $r$, this pair ``10"
will inevitably occur at binary positions not included among the
(at most $p$) qubits representing the $a_i$. Thus it is not
possible to choose a fixed number $p$ so that $a_1$ and $a_2= a_1
+r $ are both expressible using only a bounded number of binary
digits. Thus the full arithmetic progression state will not be
$p$-blocked in general.

Therefore for general values of $N$, the number we wish to factor,
the state of the computer at this stage in the calculation is not
$p$-blocked.  Of course, as we have seen, certain carefully chosen
truncated arithmetic progression states may be blocked (e.g. it is
not difficult to construct 2-blocked arithmetic progressions of
length $2^m$ for any $m$), but these are highly atypical.

Actually our argument above shows more generally, that almost all
states of the form $\ket{a}+\ket{a+r} + \ket{b_1} + \ldots
+\ket{b_m}$ where $b_i > a+r$ for all $i$, will not be
$p$-blocked. Now Simon's algorithm \cite{simon,jozsa2} involves
states of the form $\ket{x_0}+\ket{x_1}$ where $x_0$ and $x_1$ are
general $n$-qubit strings. Hence this algorithm too utilises
multi-partite entanglement of unboundedly many qubits.

\section{Computations with mixed states}\label{mixed}

The significance of entanglement for pure state computations
derives from the fact that unentangled pure states (or more
generally $p$-blocked pure states) of $n$ qubits have a
description involving only $\poly (n)$ parameters (in contrast to
$O(2^n)$ parameters for a general pure state) and consequently if
entanglement is absent in a quantum computational process then the
process can be efficiently classically simulated.

The corresponding situation for mixed states is dramatically
different: we define a mixed state $\rho$ to be unentangled if it is
separable i.e. among all the possible ways of representing $\rho$ as
a mixture of pure states, there is a mixture involving only product
states. Then it may be shown \cite{zhls,vt,bcjlps} that unentangled
mixed states have a non-zero volume in the space of all mixed states.
As an explicit example \cite{bcjlps} the $n$ qubit state \beq
\label{sepball} \rho = (1-\epsilon ) \frac{1}{2^n} I + \epsilon \xi
\eeq is unentangled for all mixed states $\xi$ if $\epsilon <
\frac{1}{4^n}$. Hence unentangled mixed states require the same
number of parameters for their description as do general mixed
states. Consequently if a quantum algorithm has unentangled (mixed)
states at each stage then the classical simulation by direct
computation of the updated state, fails to be efficient. From this
parameter counting point of view an unentangled mixed state has the
same capacity for coding information as a general mixed state so it
is plausible that the computational power of general mixed (or pure)
quantum states is already fully present in the restricted case of
unentangled mixed states. Stated otherwise, we have the following
fundamental (unresolved) question. Suppose that a quantum computation
of $N$ steps on mixed states has a {\em separable} state $\alpha_j$
at each stage $j$. Suppose also that the starting state $\alpha_0$
has a $\poly(N)$ sized description. Then, can the algorithm be
classically efficiently simulated?

At first sight one might expect that the direct computational
simulation method used in the previous section could be adapted to
become efficient. Indeed a separable state is just a classical
probabilistic mixture of unentangled pure states and we {\em can}
efficiently simulate unentangled pure state processes so maybe we
could just supplement the latter with some extra classical
probabilistic choices. Indeed such a modification works in the
restricted case of {\em classical} probabilistic processes i.e.
processes in which the state at any stage is a probabilistic mixture
of computational basis states. As a simple example consider a process
involving $n$ coins and the $j^{\rm th}$ step is to toss the $j^{\rm
th}$ coin. Then the complete state description grows exponentially
with $j$ (being a probability distribution over $2^j$ outcomes at
state $j$). But to simulate the process (i.e. sample the final
distribution once) we do not need to carry along the entire
distribution -- we just make probabilistic choices along the way
(i.e. follow one path through the exponentially large branching tree
of possibilities, rather than computing the whole tree and then
sampling the final total distribution at the end).

Unfortunately this idea appears not to work in the quantum case.
Suppose that $\rho_1$ is an unentangled state, being a mixture of
product states $\ket{\xi_i}$ with probabilities $p_i$ i.e. $\rho_1 =
\sum p_i \proj{\xi_i}$. Suppose that $\rho_2 = U\rho_1 U^\dagger$ is
also unentangled for a unitary operation $U$ (the computational
step). Let $\ket{\eta_i} = U\ket{\xi}$. Then $\rho_2$ is a mixture of
state $\ket{\eta_i}$ with probabilities $p_i$ but there is no
guarantee that $\ket{\eta_i}$ are product states! As a simple example
with 2 qubits let $\rho_1$ be an equal mixture of the product states
$\ket{\xi_1}=(\ket{0}+\ket{1}) \ket{1}$ and
$\ket{\xi_2}=(\ket{0}-\ket{1}) \ket{1}$ (where we omit normalisation
factors) and let $U$ be the controlled NOT gate. Then by updating
these pure state we get $\rho_2$ as an equal mixture of two maximally
entangled states $\ket{\eta_1} = \ket{0}\ket{1}+\ket{1}\ket{0}$ and
$\ket{\eta_2} = \ket{0}\ket{1}-\ket{1}\ket{0}$. Although these
component states are entangled, the {\em overall} mixture is
unentangled, being equivalent to an equal mixture of the product
states $\ket{0}\ket{1}$ and $\ket{1}\ket{0}$. The existence of a
separable mixture is not a property of individual component states
$\ket{\eta_1}$ but a {\em global} property of the whole ensemble $\{
\ket{\eta_i}; p_i \}$. Hence we cannot follow a single probabilistic
path of pure states through a branching tree of possibilities since
these pure states become entangled (and so the simulation becomes
inefficient). At each stage we need to re-compute a new {\em
separable} decomposition of the state. This computation generally
requires knowledge of the whole ensemble and hence cannot be
efficient in the required sense.

Quantum computation with mixed states has attracted much attention
in recent years because of the experimental implementation of
quantum computation using liquid state NMR techniques \cite{NC}
which utilises an interesting class of mixed states. The basic
idea is to consider so-called pseudo-pure states of the form \beq
\label{pspure} \rho = (1-\epsilon ) \frac{1}{2^n} I + \epsilon
\proj{\psi} \eeq (which occur in NMR experiments for suitably
small $\epsilon$). Then for any unitary transformation $U$ the
state $U\rho U^\dagger$ is also pseudo-pure with $\ket{\psi}$
replaced by $U\ket{\psi}$. Hence given any pure state quantum
algorithm, if we implement it on the pseudo-pure analogue of its
starting state, the entire pure state algorithm will unfold as
usual on the pure state perturbation $\epsilon \proj{\psi}$ of eq.
(\ref{pspure}). Furthermore if $A$ is any traceless observable
then from eq. (\ref{pspure}) we get the expectation value \beq
\label{expval} \langle A \rangle = \tr A\rho = \epsilon \bra{\psi}
A\ket{\psi} \eeq i.e. we obtain the average value of $A$ in the
pure state $\ket{\psi}$ but the signal is attenuated by
$\epsilon$.

We have seen in eq.(\ref{sepball}) that for sufficiently small
$\epsilon$ all pseudo-pure states are separable so we have the
intriguing possibility of implementing any quantum pure state
algorithm with its original running time, in a setting with no
entanglement at all! Can this provide a computational benefit
(over classical computations) in the total absence of
entanglement? Although this not not known for general algorithms,
it has been shown for Shor's algorithm (and structurally related
algorithms) \cite{lindpop} and for Grover's algorithm \cite{braun}
that the answer is negative: if the pseudo-pure states are {\em
required} to remain separable at each stage in these algorithms
then it can be proven \cite{lindpop,braun} that the value of
$\epsilon$ must decrease exponentially with input size.
Consequently to obtain the output result reliably via an
expectation value as in eq. (\ref{expval}) we either need to
repeat the algorithm exponentially many times or else, use an
exponentially increasing bulk of fluid in the liquid state NMR
implementation. In either case the implementation becomes
inefficient. Indeed equivalent possibilities are available in a
purely classical setting. For example a classical algorithm for
factoring $N$ that test divides by all numbers up to $\sqrt{N}$
can be run in polynomial time if we have an exponential bulk of
parallel computers available, or else in exponential time on a
single computer if we run the trial divisions sequentially.

\section{Is entanglement a key resource for computational
power?}\label{what?}

Recall that the significance of entanglement for pure state
computations derives from the fact that unentangled pure states (or
more generally $p$-blocked pure states) of $n$ qubits have a
description involving only $\poly (n)$ parameters (in contrast to
$O(2^n)$ parameters for a general pure state). But this special
property of unentangled states (of having a ``small'' descriptions)
is contingent on a particular mathematical description, as amplitudes
in the computational basis. If we were to adopt some other choice of
mathematical description for quantum states (and their evolution)
then although it will be mathematically equivalent to the amplitude
description, there will be a different class of states which now have
a polynomially sized description i.e. two formulations of a theory
which are mathematically equivalent (and hence equally logically
valid) need not have their corresponding mathematical descriptions of
elements of the theory being interconvertible by a {\em polynomially
bounded} computation. With this in mind we see that the significance
of entanglement as a resource for quantum computation is not an {\em
intrinsic} property of quantum physics {\em itself} but is tied to a
particular additional (arbitrary) choice of mathematical formalism
for the theory.

Thus suppose that instead of the amplitude description we choose
some other mathematical description $\cd$ of quantum states (and
gates). Indeed there is a rich variety of possible alternative
descriptions. Then there will be an associated property $prop (\cd
)$ of states which guarantees that the $\cd $-description of the
quantum computational process grows only polynomially with the
number of qubits if $prop (\cd )$ is absent (e.g. if $\cd$ is the
amplitude description then $prop (\cd )$ is just the notion of
entanglement). Thus, just as for entanglement, we can equally well
claim that $prop (\cd )$ for {\em any} $\cd$, is an essential
resource for quantum computational speed-up! Entanglement itself
appears to have no special status here.

An explicit example of an alternative formalism and its
implications for the power of quantum computation is provided by
the so-called stabiliser formalism and the Gottesman--Knill
theorem \cite{NC,gotth}. The essential ingredients are as follows.
(See the previous references for details and proofs). The Pauli
group $\cp_n$ on $n$ qubits is the group generated by all $n$-fold
tensor products of of the Pauli matrices $\sigma_x, \sigma_y,
\sigma_z$, the 1-qubit identity operator $I$ and multiplicative
factors of -1  and $i$. Any subgroup $\ck$ of $\cp_n$ may be
described by a list of at most $n$ elements which generate the
subgroup, so any subgroup has a $\poly (n)$ sized description. We
write $\ck = [ g_1 , \ldots ,g_k ]$ if $g_1,\ldots ,g_k$ is a set
of generators for $\ck$. For each $n$ some states $\ket{\alpha}$
of $n$ qubits have a special property that they are stabilised by
a subgroup $\ck_\alpha = [ g_1 , \ldots ,g_k ]$ ($k\leq n$) of
$\cp_n$ i.e. $\ket{\alpha}$ is the unique state such that $g_i
\ket{\alpha}=\ket{\alpha}$ for $i=1, \ldots ,k$. For example
$\ket{0}\ket{0}+\ket{1}\ket{1}$ is the unique state stabilised by
$[\sigma_x \otimes \sigma_x , \sigma_z\otimes \sigma_z ]$.

Let $\cs$ be the class of all such states (for all $n$). States in
$\cs$ can be mathematically specified by giving a list of
generators of the corresponding stabiliser subgroup i.e. states in
$\cs$ have a $\poly (n)$ sized stabiliser description. Then we
have the following facts:
\\ (a) All computational basis states are in $\cs$; \\ (b) The
following gates all preserve $\cs$ and have simple (i.e.
efficient) update rules for their effect on the stabiliser
description of the states: \[ \mbox{the 1-qubit gates (on any
qubit):}\hspace{6mm}  \left( \begin{array}{cc} 1 & 0 \\ 0 & i
\end{array} \right),\,\, H = \frac{1}{\sqrt{2}} \left( \begin{array}{cc} 1 &
1
\\ 1 & -1
\end{array} \right),\,\,  \sigma_x,\,\, \sigma_y,\,\,
\sigma_z \] and the 2-qubit controlled-NOT gate (on any two
qubits);\\ (c) Outcome probabilities for a measurement in the
computational basis are efficiently computable from the stabiliser
description of the state; \\ (d) Application of other gates (such as
the Toffoli gate or $\frac{\pi}{8}$ phase gate \cite{NC} which would
provide universal computation with the gates in (b)) will generally
transform states $\ket{\alpha}$ in $\cs$ to states $\ket{\psi}$
outside of $\cs$. Recall that any unitary transformation may be
expressed as a linear combination of $I,\sigma_x, \sigma_y$ and
$\sigma_z$ so we could introduce a stabiliser description of
$\ket{\psi}$ as a suitable subalgebra of the group algebra of
$\cp_n$. But this description of $n$-qubit states $\ket{\psi}$ would
not generally remain $\poly (n)$ sized if a computational process
involves such more general gates.

From (a), (b) and (c) we immediately get: \\ {\em Gottesman--Knill
theorem:} Any quantum computational process that starts with a
computational basis state and uses only the gates in (b) above (so
that the states remains in $\cs$ at each stage) can be efficiently
classically simulated (by computation in the stabiliser description).

Note that such computations can generate large amounts of
multi-partite entanglement of unboundedly many parties (e.g. by
repeated use of $H$ and controlled-NOT) so that in contrast to the
stabiliser formalism, if we use the amplitude formalism then the
computation will have an exponentially growing description.

Thus if $prop (\cs)$ denotes the property of a state that it does
not have a polynomially sized stabiliser description, then we can
claim that $prop(\cs)$ is an essential resource for quantum
computational power (since absence of $prop(\cs)$ implies
efficient classical simulability).

The concept of the stabiliser description of a state (compared to
the amplitude description) provides a hint of how conceptually
diverse alternative formulations of quantum theory can be. Some
recent work by Valiant \cite{valiant} and Terhal and DiVincenzo
\cite{terdiv} identifying another interesting class of quantum
computations that can be efficiently simulated, appears to also
fit into this framework, utilising a fermionic operator formalism
as a mathematical description of quantum computational processes.

Thus in a fundamental sense, the power of quantum computation over
classical computation ought to derive simultaneously from {\em all}
possible classical mathematical formalisms for representing quantum
theory, not any single such formalism and associated quality (such as
entanglement) i.e. we have arrived at the enigmatic prospect of
needing a representation of quantum physics that does not single out
any particular choice of mathematical formalism.

\bigskip
\noindent {\Large\bf Acknowledgements}

RJ is supported by the U.K. Engineering and Physical Sciences
Research Council. This work was initiated in 1999 in collaborative
opportunities provided by the Workshop on Quantum Information,
Computation and Decoherence, at the Isaac Newton Institute for
Mathematical Sciences, Cambridge, UK. RJ and NL also acknowledge the
support of the European 5$^{\rm th}$ framework networks QAIP
(Contract IST-1999-11234) and EQUIP (Contract IST-1999-11063).

\end{document}